\documentclass[journal=apchd5,manuscript=article]{achemso}
\setkeys{acs}{doi=true}

\newcommand{\doi}[1]{\href{https://doi.org/#1}{\nolinkurl{#1}}}

\usepackage[section]{placeins}

\usepackage{amsmath}

\usepackage{amssymb}
\usepackage{contour}
\usepackage{dsfont}
\usepackage{bm}
\usepackage{graphicx}

\contournumber{40}
\contourlength{1pt}

\usepackage{amsmath,amsfonts,amssymb}
\usepackage{graphicx,subcaption,multirow}
\usepackage[colorlinks=false,bookmarks=false,allcolors=black]{hyperref}
\hypersetup{pdfstartview={FitH},pdfpagemode={UseNone}, breaklinks=true, colorlinks,allcolors=black, bookmarksopen=true, pdfnewwindow=true}
\usepackage[all]{hypcap}

\graphicspath{{pict/}{figures/}{}}

\usepackage[english]{babel}

\title{Robust Classical and Quantum Polarimetry with a Single Nanostructured Metagrating}

\author{Shaun~Lung}
\email{shaun.lung@uni-jena.de}
\affiliation{Abbe Center of Photonics, Friedrich-Schiller Universit{\"a}t, Albert-Einstein-Stra{\ss}e 15, 07745 Jena, Germany}
\alsoaffiliation{ARC Centre of Excellence for Transformative Meta-Optical Systems (TMOS), Department of Electronic Materials Engineering, Research School of Physics, The Australian National University, Canberra, ACT~2600, Australia}
\author{Kai~Wang} 
\affiliation{Department of Physics, McGill University, 3600 rue University, Montreal, Quebec H3A 2T8, Canada}
\author{Nicolas~R.H.~Pedersen}
\author{Frank~Setzpfandt}
\affiliation{Abbe Center of Photonics, Friedrich-Schiller Universit{\"a}t, Albert-Einstein-Stra{\ss}e 15, 07745 Jena, Germany}
\alsoaffiliation{Fraunhofer Institute for Applied Optics and Precision Engineering, 07745 Jena, Germany}
\author{Andrey~A.~Sukhorukov}
\affiliation{ARC Centre of Excellence for Transformative Meta-Optical Systems (TMOS), Department of Electronic Materials Engineering, Research School of Physics, The Australian National University, Canberra, ACT~2600, Australia}


\date{{\footnotesize \today}}

\begin{document}
\sloppy
\maketitle

\begin{abstract} 
We formulate a new conceptual approach for one-shot complete polarization state measurement with nanostructured metasurfaces applicable to classical light and multi-photon quantum states, by drawing on the principles of generalized quantum measurements based on positive operator-valued measures (POVMs). Accurate polarization reconstruction from a combination of photon counts or correlations from several diffraction orders is robust with respect to even strong fabrication inaccuracies, requiring only a single classical calibration of metasurface transmission.
Furthermore, this approach operates with a single metagrating without interleaving, allowing for the metasurface size reduction while preserving the high transmission efficiency and output beam quality.
We theoretically obtained original metasurface designs, fabricated the metasurface from amorphous silicon nanostructures deposited on glass, and experimentally confirmed accurate polarization reconstruction for laser beams. 
We also anticipate robust operation under changes in environmental conditions, opening new possibilities for space-based imaging and satellite optics.
\end{abstract}

\section*{Introduction}
Single-shot optical polarimetry using ultra-thin nanostructured metasurfaces opens up new opportunities for diverse applications~\cite{Martinez:2018-750:SCI, Intaravanne:2020-1003:NANP, Rubin:2021-836:ADOP}, facilitating the measurement of both classical~\cite{Pors:2015-716:OPT, Zhang:2019-1190:OPT, Arbabi:2018-3132:ACSP, Wei:2017-1580:OL, Yang:2018-4607:NCOM} and quantum~\cite{Wang:2018-1104:SCI} polarization states.
Whereas traditional polarimetry involves multiple measurements performed via physically varying bulk optical elements such as waveplates and polarizers~\cite{Chekhova:2021:PolarizationLight},
single-shot approaches remove the need for 
reconfigurability, thereby avoiding the associated measurement errors and facilitating real-time polarization state monitoring~\cite{Rubin:2019-eaax1839:SCI, Rubin:2022-9389:OE} that can be combined with spectral imaging~\cite{Chen:2016-224002:NANT, Ding:2017-943:ACSP, Sun:2022-2100650:LPR, Camayd-Munoz:2020-280:OPT}.

Similar to conventional classical~\cite{Chekhova:2021:PolarizationLight} and quantum polarimetry~\cite{James:2001-52312:PRA}, metasurfaces were initially designed to perform several complimentary projection measurements~\cite{Foreman:2015-263901:PRL}, such that each of the outputs corresponds to a specific polarization~\cite{Pors:2015-716:OPT, Martinez:2018-750:SCI, Intaravanne:2020-1003:NANP, Rubin:2021-836:ADOP}.
This functionality can be accomplished with metasurfaces that split particular polarization components into distinct diffraction orders or focal spots.
For the tomographic characterization of quantum states, metasurfaces were previously developed by interleaving several metagratings whose number scales linearly with the number of photons~\cite{Wang:2018-1104:SCI}.
However, the simple interleaving (i) limits the device compactness since the photons have to spatially overlap with multiple gratings at once and (ii) introduces output beam distortions which reduce the detection efficiency. This poses a question on how to overcome such limitations and thereby support the emerging applications of metasurfaces in quantum imaging~\cite{Pittman:1995-3429:PRA, Altuzarra:2019-20101:PRA, Vega:2021-64032:PRAP}.

We reveal that efficient polarimetry with metasurfaces can be accomplished without the commonly considered requirement of realizing close-to-perfect polarization projection measurements. 
Remarkably, even if each of the metasurface output ports represents a partial polarizer operation that by itself provides inconclusive information on the input state, a tailored combination of all outputs allows for very accurate polarization reconstruction. We achieve this by adopting the framework of generalized quantum measurements based on positive operator-valued measures (POVMs)~\cite{Brandt:1999-434:AMJP, Hamieh:2004-52325:PRA, Renes:2004-2171:JMP, Ziman:2008-62112:PRA, Shapiro:2008-52330:PRA, AlKhafaji:2022-458352:OE, Martinez:2023-190:NPHYS} for the metasurface design.
Such an approach fundamentally improves the robustness with respect to nanofabrication inaccuracies and also extends the flexibility in metasurface designs, allowing in particular for higher efficiency and output beam shaping with small-area metasurfaces. We present simulation results for one- and two-photon states. Then, we experimentally demonstrate the operation with laser light that illustrates the classical regime and also emulates a single-photon quantum case~\cite{Barnett:2022-114004:PS}.










\begin{figure}
    \centering
    \includegraphics[width=\textwidth]{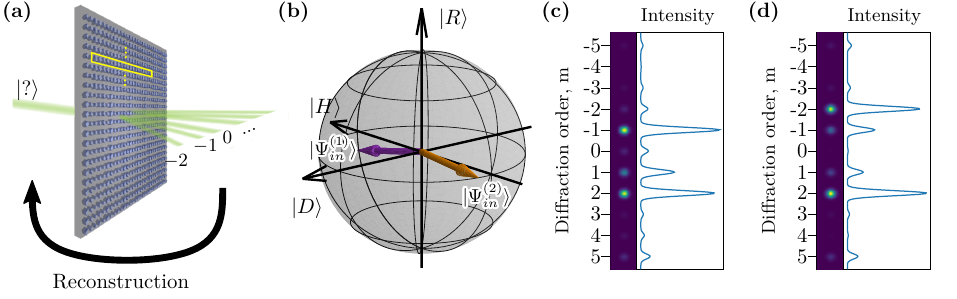}
    \vspace{0.5em}
    \caption{\label{fig:theory_and_design}
        (a)~Conceptual sketch of single-metagrating polarimetry. Highlighted in yellow is a single 16 resonator unit cell. From the output diffraction orders, any unknown input state may be reconstructed by knowing the metasurface instrument matrix.
        (b)~Poincar\'e sphere representation of two arbitrarily chosen input states $|\Psi_{in}^{(1)}\rangle$, $|\Psi_{in}^{(2)}\rangle$ characterized in Jones formalism by the polarization angles and phases $(1.5, 0.1)$, $(0.1, 1.5)$ respectively.
        (c,d)~Intensities of output diffraction orders corresponding to the respective input states $|\Psi_{in}^{(1)}\rangle$, $|\Psi_{in}^{(2)}\rangle$.
    }
\end{figure}

\section*{Theory of multi-photon polarization measurements} \label{sec:theory}

We first formulate a general theory of polarization measurements. We consider a metasurface, which splits an input beam into several diffraction orders as sketched in Fig.~\ref{fig:theory_and_design}(a). Then, the measurements of photon correlations between the output ports can be used to reconstruct the input quantum state. In a previous study~\cite{Wang:2018-1104:SCI}, it was assumed that the photon polarization is specifically selected at each diffraction order. This effectively required the metasurface to act as a near-perfect polarization splitter, which came at a cost of larger device area requirements due to a need to interleave several gratings and the distortion of the output beam profile. In the following, we show that such limitations can be removed, allowing for a compact and robust metasurface design.

Since we consider a linear regime, the transformation of quantum states can be expressed through the classical Jones transfer matrices $\mathbf{T}^{(m)}$ to each of the diffraction orders number $m$,
%
%
\begin{equation}\label{eq:psi_out_jones}
    \psi_{out}^{(m)} = \mathbf{T}^{(m)} \psi_{in} \,,
\end{equation}
where $\psi_{in}$ and $\psi_{out}^{(m)}$ are the classical input and output polarization states in the form of Jones vectors at the respective diffraction orders.
Each diffraction order functions, in general, as a partial polarizer acting on the incoming light. It is then convenient for the following analysis to perform a singular value decomposition (SVD) of the transfer matrices,
\begin{equation} \label{eq:TmSVD}
  \mathbf{T}^{(m)} = 
  \textbf{U}^{(m)}\, 
  \begin{pmatrix}
      \zeta_{m,1} & 0 \\
      0 & \zeta_{m,2}
  \end{pmatrix}
  \left(\textbf{W}^{(m)}\right)^\dagger \,,
\end{equation}
where $\zeta_{m,1}, \zeta_{m,2}\ge 0$ are the singular values,  $\textbf{U}^{(m)}=[\textbf{U}^{(m)}_1,\textbf{U}^{(m)}_2]$ and $\textbf{W}^{(m)}=[\textbf{W}^{(m)}_1,\textbf{W}^{(m)}_2]$ are unitary matrices, the subscripts indexing the respective columns. We choose the order of singular values such that $\zeta_{m,1} \ge \zeta_{m,2}$. 
%
If $\zeta_{m,2}=0$, the polarization state at the corresponding diffraction order is fixed as for a case of a perfect polarizer. However, in a general case of $\zeta_{m,2}>0$, the metasurface effectively acts as a partial polarizer, with a power extinction ratio of $(\zeta_{m,2}/\zeta_{m,1})^2$.

We now formulate the transformation by the metasurface of the photon creation and annihilation operators from the input,  
$\hat{a}_p^\dagger$ and $\hat{a}_p$, where $p=\{H,V\}$ is the polarization state, to each of the output diffraction orders, $\hat{b}_{m,p}^\dagger$ and $\hat{b}_{m,p}$. This can be expressed through the linear transfer matrix elements as:
\begin{equation} \label{eq:operators}
   \hat{b}_{m,p} = \sum_{p'=H,V} \left(\mathbf{T}^{(m)}_{p,p'}\right)^{\ast}\, \hat{a}_{p'} \;, \quad
   \hat{b}_{m,p}^\dagger = \sum_{p'=H,V} \mathbf{T}^{(m)}_{p,p'}\, \hat{a}_{p'}^\dagger \;.
\end{equation}

We target the tomography of quantum polarization-entangled states with a fixed photon number $N$ at the input, which is a common practical task~\cite{James:2001-52312:PRA, Titchener:2016-4079:OL, Oren:2017-993:OPT, Titchener:2018-19:NPJQI},
considering the simplest type of click detectors that cannot resolve the number of arriving photons and cannot distinguish the photon polarization state.
Notably, the state characterization can also be generalized to the regime when the maximum photon number is known using the approach of Ref.~\cite{Bayraktar:2016-20105:PRA}.
If no more than one photon arrives at such a detector positioned at the diffraction order $m$, then its response is governed by the following POVM operator \cite{Kok:2010:OpticalQuantum},
\begin{equation} \label{eq:Am}
   \hat{A}^{(m)} = \sum_{p=H,V} \hat{b}_{m,p}^\dagger \hat{b}_{m,p} = \sum_{p,p'=H,V} \textbf{A}^{(m)}_{p,p'}\, \hat{a}_{p}^\dagger \hat{a}_{p'}\;,
\end{equation}
where using Eqs.~(\ref{eq:TmSVD}) and~(\ref{eq:operators}) we calculate the matrix expression
\begin{equation}
    \label{eq:projectors}
    \mathbf{A}^{(m)} = \left(\mathbf{T}^{(m)}\right)^\dagger \mathbf{T}^{(m)}
    = \sum_{j=1,2} \zeta_{m,j}^2 \left|\textbf{W}^{(m)}_j\right\rangle \left\langle \textbf{W}^{(m)}_j \right|\, 
    = \left(\zeta_{m,1}^2 - \zeta_{m,2}^2\right) \left|\textbf{W}^{(m)}_1\right\rangle \left\langle \textbf{W}^{(m)}_1 \right| + \zeta_{m,2}^2 \textbf{I} \, 
    .
\end{equation}
We see that this is a sum of a polarization projection operator and a polarization-insensitive detection. The presence of the latter term is a consequence of the partial-polarizer transformation at each of the diffraction orders. Although conventional polarimetry requires near-perfect polarizers (i.e. $\zeta_{m,2}=0$), we apply the POVM formalism that enables unique and accurate quantum state reconstruction in the regime of $\zeta_{m,2}>0$.
Beyond metasurfaces, we expect that the formulated approach can be also used to enhance polarization measurements using nanowire detectors~\cite{Zhu:2021-2105729:ADM}, overcoming the limitations due to relatively low polarization extinction ratios~\cite{Park:2015-7209:OE}.

After determining the detection operators, we find the probabilities of the simultaneous detection of $N$ photons by a combination of $N$ detectors at the diffraction orders $m_1, m_2, \ldots, m_N$, when there is exactly one photon at each detector. Notably, if more than one photon arrives at a particular detector, then the total number of coincidences measured across all detectors will be less than $N$, and thus excluded from the analysis. 
The $N$-detector correlations can be calculated as 
\begin{equation} \label{eq:correl}
  \Gamma(m_1, m_2, \ldots, m_N) = {\rm Tr}\left( \rho^{(N)} \hat{A}_{m_1} \hat{A}_{m_2} \cdots \hat{A}_{m_N} \right) .
\end{equation}
Here, $\rho^{(N)}$ is an input density matrix.
Then, we follow an established procedure~\cite{Wang:2018-1104:SCI, Wang:2019-41:OPT} to enumerate with index $q$ all the possible $N$ combinations of $M$ detectors,
$(m_1,m_2,\ldots,m_N)$, and rewrite Eq.~(\ref{eq:correl}) in an equivalent form,
\begin{equation} \label{eq:correlMatrix}
  \Gamma_q = \sum_{s=1}^S {\bf B}_{q,s}  r_s \;.
\end{equation}
Here $r_s$ are the independent real and imaginary parts of the input density matrix defined according to the procedure in Ref.~\cite{Titchener:2018-19:NPJQI}, $S = (N+3)!/(3! N!)$, $q=1,\ldots,Q$, $Q = M!/(N!(M-N)!)$, and $M$ is the total number of detected diffraction orders.
The matrix elements ${\bf B}_{p,s}$ depend on the transfer matrix elements, and more specifically on the vectors $\textbf{W}^{(m)}_j$ and singular values $\zeta_{m,p}$ according to the form of Eq.~(\ref{eq:projectors}).

We can then reconstruct an input state from the correlation measurements by performing a pseudo-inversion of Eq.~(\ref{eq:correlMatrix}), provided the number of different correlations matches or exceeds the number of unknowns, $Q \ge S$, which is satisfied when
\begin{equation} \label{eq:necessaryCondition}
    M \ge N+3 \; .
\end{equation}
%

Importantly, in addition to the necessary condition in Eq.~(\ref{eq:necessaryCondition}), it is essential that reconstruction results are robust in presence of experimental errors in the correlation measurements~\cite{Filippov:2010-32:JRLR}. This can be expressed as a requirement to minimize the condition number $\kappa$ of matrix ${\bf B}$, defined as a ratio of its largest and smallest singular values~\cite{Bogdanov:2010-686:JETPL, Miranowicz:2014-62123:PRA, Wang:2018-1104:SCI}.
The condition number of an equation characterizes the worst-case error in the output for a given error in the set of input parameters. In particular, the condition number of a linear Eq.~(\ref{eq:correlMatrix}) characterizes the accuracy of the reconstruction against errors in the observations.

\section*{Experimental methods}

\subsection*{Optimized metagrating design} \label{sec:optimize}

\begin{figure}
    \centering
    \includegraphics[width=\textwidth]{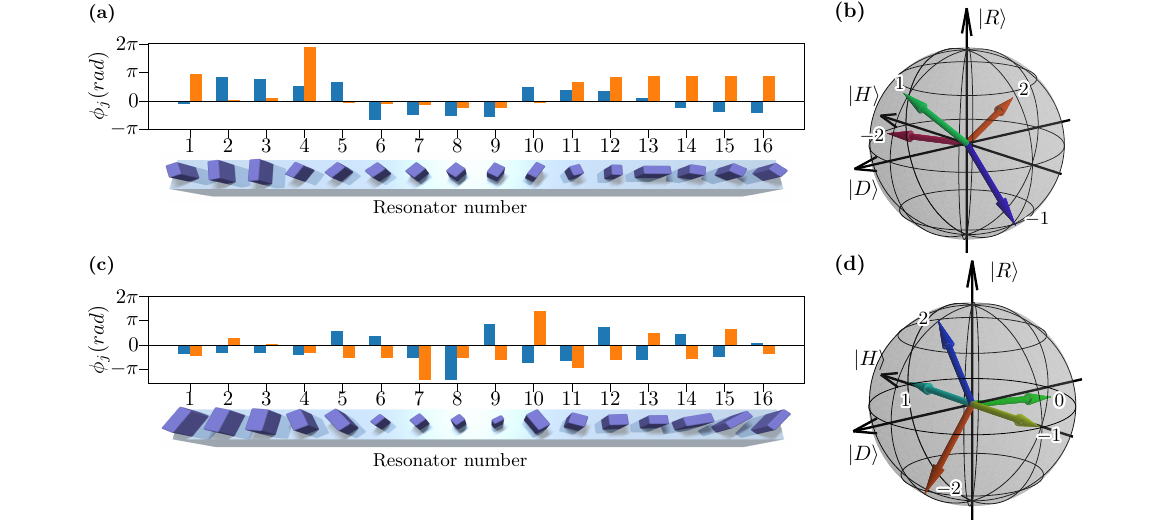}
    \vspace{0.5em}
    \caption{\label{fig:parameters_and_bases}
    Design of metagratings for (a,b)~one-photon or classical and (c,d)~two-photon polarimetry.
    (a,c)~Bottom - rendering of metagratings made of amorphous silicon on glass substrate, illustrating different shapes and orientations of rectangular nanopillars.
    Top - the phase retardances along the two principal axes of nanopillars shown with blue and orange colors.
    (b,d)~Corresponding Poincar\'e sphere representation of partial polarizer transformations at different diffraction orders as labeled.
    }
\end{figure}

To implement the multiple polarization transformations required to implement the proposed polarimetry scheme, we leverage the flexibility of designing metasurfaces as follows. 
First, we defined a metagrating consisting of $L$ rectangular nanoresonators arranged into a periodic supercell, such as shown in Figs.~\ref{fig:parameters_and_bases}(a,c).
A transfer matrix of each nanoresonator can be represented in the Jones formalism~\cite{Arbabi:2015-937:NNANO} as
\begin{equation}
    \mathbf{U}^{(l)}=\mathrm{\mathbf{R}}\left(-\theta^{(l)}\right)
    \cdot
    \begin{bmatrix}
        e^{i\phi_o^{(l)}} & 0 \\
        0 & e^{i\phi_e^{(l)}}
    \end{bmatrix}
    \cdot
    \mathrm{\mathbf{R}}\left(\theta^{(l)}\right) \,,
    \label{eq:jones_cuboid}
\end{equation}
where $l$ is the resonator number, $\phi_o^{(l)}$ and $\phi_e^{(l)}$ are the phase shifts imposed by the resonator along the ordinary and extraordinary axes, and $\mathrm{\mathbf{R}}\left(\theta^{(l)}\right)$ is a two-by-two rotation matrix by angle $\theta^{(l)}$.
Then, we calculate the transfer matrices of the metasurface at the output diffraction orders using the Fourier transform,
\begin{equation} \label{eq:fourier}
    \mathbf{T}^{(m)} = \sum_{l=0}^L\mathbf{U}^{(l)} e^{-\frac{2\pi i}{ N}m l} \,.
\end{equation}
%
We calculate these matrices numerically and then use Eqs.~(\ref{eq:Am})-(\ref{eq:correlMatrix}) to determine the corresponding condition number $\kappa$, defined as the ratio of the maximum and minimum  singular value decomposition values of matrix ${\bf B}$ appearing in Eq.~(\ref{eq:correlMatrix}). By minimizing the condition number via a gradient-descent algorithm, it is thus possible to obtain a highly robust design. 
%
This optimization produces a set of phase parameters corresponding to each pixel of the unit cell, consisting of phase shifts along the ordinary and extraordinary axes as well as the orientation angles.

We present the optimized metagrating designs for one- and two-photon cases in Figs.~\ref{fig:parameters_and_bases}(a,c) for 
the minimum required number of outputs according to
Eq.~(\ref{eq:necessaryCondition}), $M=4$ and $M=5$, respectively. The numerical values of the nanoresonator parameters are presented in Supplementary Information Sec.~S1.
In these examples, we chose to consider an array of $L=16$ nanoresonators, but it should be observed that this choice is neither unique nor constrained, and different numbers can be selected depending on the required angular diversion of the diffraction orders after the metasurface.


As implied by Eq.~(\ref{eq:projectors}), the vectors $\mathbf{W}_j^{(m)}$ are the basis states of the polarization measurement. After converting them to a Stokes basis, we plot them on a Poincar\'e sphere for convenient visualization of the basis states at different diffraction orders for a given metasurface design, as shown in Figs.~\ref{fig:parameters_and_bases}(b,d). We define the vector lengths as 
\begin{equation}
    \label{eq:pol_basis_length}
    R = 1 - \frac{\zeta_{m,2}^2}{\zeta_{m,1}^2},
\end{equation}
where 
$\zeta_{m,2}^2$ and $\zeta_{m,1}^2$ are by definition proportional to
the minimum and maximum powers transmitted to the relevant diffraction order $m$ across a set of all possible polarization states under the same input power. We note that $R=1$ corresponds to a fully polarized output, as would be required for projective measurements. We observe that $R<1$ for several outputs of our optimized metagratings, meaning that they act as partial polarizers with the finite extinction ratio $(1-R)^{-1}$. Nevertheless, this allows for efficient reconstruction of the input polarization states. 

Indeed, the corresponding inverse condition numbers, calculated for the metagratings that are numerically optimized to maximize their values, are around $0.37$ and $0.21$ in the spectral region of interest for single- and two-photon states. The single-photon value is comparable to the theoretically-best maximum~\cite{Foreman:2015-263901:PRL} of $1/\sqrt{3} \simeq 0.58$. While the best possible value for two-photon state reconstruction with click detectors, which do not resolve the events of both photons exiting from the same port, is not known analytically, previous numerical optimisations of integrated waveguide circuits~\cite{Titchener:2016-4079:OL, Wang:2019-41:OPT} reported values up to $\sim 0.25$, again close to the two-photon result above in our current study.



\subsection*{Robustness to fabrication errors}

\begin{figure}
    \centering
    \includegraphics[width=0.7\textwidth]{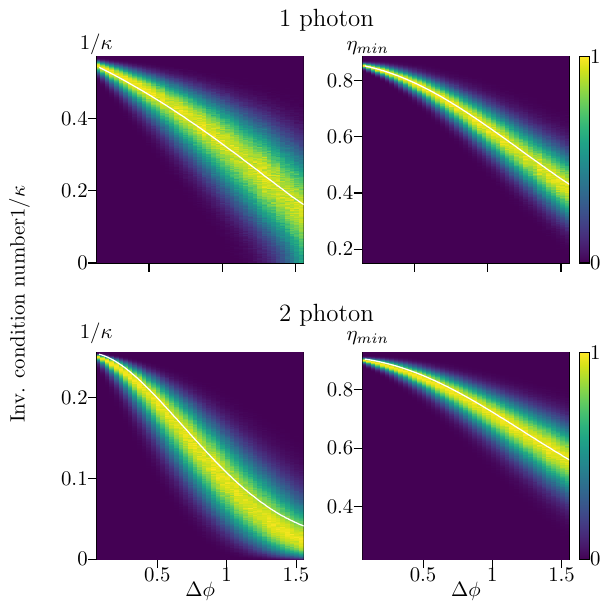}
    \vspace{0.5em}
    \caption{\label{fig:robustness}
        Simulated effects of random phase errors, up to a maximum of $\Delta\phi$, applied to each of the nanoresonators in the metagratings optimized for one-photon and two-photon polarimetry as illustrated in Figs.~\ref{fig:parameters_and_bases}(a) and (c), respectively.
        Shown are the probability densities, normalized to a maximum of unity at each $\Delta\phi$, for the inverse condition numbers $1/\kappa$ and diffraction efficiencies $\eta_{min}$. White lines indicate the average values. 
    }
\end{figure}

Based on the POVM formulation of the design, strong robustness against fabrication errors is expected of the metasurface, since accurate reconstruction is possible even if the polarization extinction ratios at individual outputs are affected. 
We have carried out numerical simulations to demonstrate this, considering as representative examples the 
metasurface designs for single- and two-photon polarimetry presented above.

Under realistic fabrication scenarios, the most common deviations from the analytical design pertain to the overall sizes of the nanoresonators. These were modelled as variance in the phase shifts along the ordinary and extraordinary axes of the individual nanopixels, corresponding to the decomposition shown in Eq.~(\ref{eq:jones_cuboid}), and thus, random errors up to $\Delta\phi$ were added to each nanopixel, and then the transmission of the altered metasurface structure was calculated via Eq.~(\ref{eq:fourier}).
From this result, we computed for the perturbed metasurface the overall inverse condition numbers $1/\kappa$ and diffraction efficiency $\eta_{min}$.

The diffraction efficiency is defined as the minimum fraction of the total input power that is diffracted to the chosen orders used in computing the corresponding inverse condition number, specifically to the ($\pm1$, $\pm2$) orders in the one- and (0, $\pm1$, $\pm2$) orders in the two-photon case.
By repeating such simulations over $10^5$ times for each value of $\Delta\phi$, we estimated the degradation of performance vs. the degree of random errors, as summarised in Fig.~\ref{fig:robustness}. 
The errors trialed range up to an extreme case of $\Delta\phi=\pi/2$, which far exceeds the possible effects of the nanofabrication errors. 
We also note that in the two-photon case, the zero order diffraction was included in the computation of both $1/\kappa$ and $\eta_{min}$, whereas the single-photon case excludes the zero order. The additional basis state provided by the zero order diffraction was found to slightly increase the achievable diffraction efficiency in the two-photon case.

We find that for realistic levels of errors, below $10^{-1}$, both the inverse condition number and the diffraction efficiencies do not drop significantly, as shown in Fig.~\ref{fig:robustness}. However, as one might expect, the two-photon design is more sensitive to errors than the single-photon case, owing to the necessary consideration of additional diffraction orders to fully resolve multiple photons.



\section*{Experimental results}

\subsection*{Metasurface patterns: design and fabrication}

We first determine the physical dimensions of nanoresonators realizing the required phase delays, such as those summarized in Supplementary Information Sec.~S1, using an established procedure~\cite{Arbabi:2015-937:NNANO, Wang:2018-1104:SCI}. Specifically, we perform a sweep of the length and width parameters of cuboidal pixels, while keeping fixed the meta-grating period, height, and refractive indices, and calculate the transmission using a numerical technique known as Rigorous Coupled Wave Analysis (RCWA)~\cite{Lalanne:1997-1592:JOSA, Lalanne:1996-779:JOSA, Hugonin:2101.00901:ARXIV}. Thereby, we produce a lookup table from which suitable designs could be simply selected. 
Thereby, one can design a metasurface for operating at the desired spectral regions.

As a demonstration, we develop a metasurface for operation at the telecommunication band, around the 1550~nm wavelength. As a platform, we choose a dielectric metasurface made of amorphous silicon on a glass substrate, leveraging the high transmissivity to ensure high efficiency of operation. 
%
The height and period used were $832nm$ and $800nm$ respectively, the former corresponding to measurements of deposited silicon that would be used to fabricate the metasurface. The physical parameters of the metasurface were thus designed by selecting pixels that fit desired phase parameters, up to an arbitrary global phase. 
The combined metasurface structure designed was then simulated using a commercial electrodynamics solver, CST Studio, as a final optimization pass and 
to check the influence of 
optical couplings between adjacent pixels that are not accounted for in the design of individual nanoresonators. This optimization pass did not alter the design singificantly, changing the design parameters by $<2\%$.

The metasurfaces were fabricated from a $832$~nm-thick amorphous silicon layer prepared at the ANU node of Australian Nanofabrication Facility (ANFF) using Plasma-Enhanced Chemical Vapor Deposition (PECVD) on a glass substrate. It was subsequently etched at the University of Jena using Electron Beam Lithography (EBL) and Inductively Coupled Plasma (ICP) etching. Slight variants in the metasurfaces were prepared by intentionally varying the EBL exposure times, allowing for a selection of best-case fabrication outcomes.

\section*{Results and discussion}\label{sec:characterization}

\begin{figure}
    \centering
    \includegraphics[width=\textwidth]{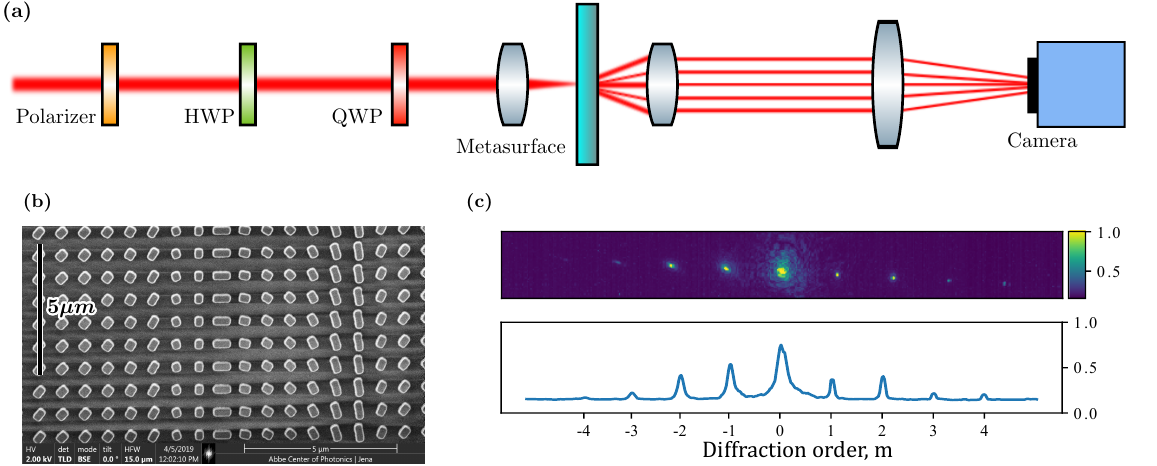}
    \vspace{0.5em}
    \caption{\label{fig:experiment}
        (a)~Schematic of the experimental setup used to classically characterize the metasurface. Input states were prepared from a variable-wavelength infrared laser using a fixed polarizer and motorized half- and quarter- waveplates, before being collimated on the metasurface by lenses. Output diffraction order intensities were collected using a CCD camera.
        (b)~Scanning Electron Microscope (SEM) image of the metasurface, fabricated as $832nm$ amorphous silicon on glass via electron beam lithography.
        (c)~Representative reading as taken using the camera, and the processed intensity plot obtained by slicing across the diffraction orders. Intensities were normalized to 1.
    }
\end{figure}

After fabrication, we measured the transmission of classical light through the metasurface.
This enables the characterization of the metasurface transformation from the input to distinct outputs, which can then be used for the reconstruction of multi-photon states. We use a classical-quantum analogy~\cite{Barnett:2022-114004:PS} since
both the single-photon (for $N=1$) counts at the outputs and the classical output intensities can be described by Eqs.~(\ref{eq:correl}),(\ref{eq:correlMatrix}). Here the density matrix can be conveniently related to the Stokes parameters $\Vec{S}$~\cite{Fano:1949-859:JOS, James:2001-52312:PRA, Goldberg:2021-1:ADOP},
\begin{equation} \label{eq:rhoStokes}
    \rho^{(1)} = \frac{1}{2} \sum_{j=0}^3 S_j {\bf \sigma}_j \,  = \frac{1}{2} \left( \begin{array}{ll} 
            S_0 + S_3 & S_1 - i S_2 \\ 
            S_1 + i S_2 & S_0 - S_3 \end{array} \right),
\end{equation}
where ${\bf \sigma}_j$ are the Pauli matrices.
For convenience, we choose the independent parameters in the input density matrix as $\Vec{r} = \Vec{S}$. Then, the matrix ${\bf B}$ in Eq.~(\ref{eq:correlMatrix}) has the meaning of an instrument matrix, since
%
\begin{equation}
    \label{eq:p_out_stokes}
       \Gamma = \mathbf{B} \, \Vec{S} 
\end{equation}
determines the intensities at the output diffraction orders for classical light with the input Stokes parameters $\mathbf{S}$. Specifically, at the output number $m$
\begin{equation}
    \label{eq:p_out_stokes_m}
       \Gamma_m = \sum_j \mathbf{B}_{m,j} \, S_j .
\end{equation}
On the other hand, using Eq.~(\ref{eq:correl}) at $N=1$ the output intensities can be expressed as
\begin{equation}
  \mathbf{\Gamma}_m = \sum_{j,l} \mathbf{A}^{(m)}_{j,l} \mathbf{\rho}^{(1)}_{l,j} \,.
\end{equation}
We then obtain 
\begin{equation} \label{eq:A_B}
    \mathbf{A}^{(m)} = 
    \left( \begin{array}{ll} 
    B_{m,1} + B_{m,4} & B_{m,2} - i B_{m,3} \\ 
    B_{m,2} + i B_{m,3} & B_{m,1} - B_{m,4} \end{array} \right) .
\end{equation}
%
In this way, the classical instrument matrix $\mathbf{B}$ can be used to determine all matrices $\mathbf{A}^{(m)}$ and thereby allow the subsequent reconstruction of arbitrary multi-photon states.

The experimental setup is schematically illustrated in Fig.~\ref{fig:experiment}(a). 
Using a half-wave plate, quarter-wave plate, and a fixed polarizer, polarization states were prepared from a variable-wavelength laser operating in the $1500-1575$~nm telecommunications bandwidth. The prepared polarization state was then focused to a spotsize of approximately $15\,\mu$m normally incident on the metasurface. The diffraction orders were then collected using an objective lens with a high numerical aperture, and imaged onto an infrared CCD camera using a convex lens. As a separate measurement, the camera was replaced with a calibrated power meter in order to determine the total power that was transmitted through the metasurface.

We measured the intensities of the diffraction orders over varying input polarization orders by capturing images on a camera, such as shown in Fig.~\ref{fig:experiment}(c). First, the diffraction order spot locations were determined,
and then the individual intensities were extracted by integrating over the area of each diffraction spot. 

We performed measurements for a calibration set of 360 distinct input polarization states and calculated the instrument matrix $\mathbf{B}$ for diffraction orders $(\pm2, \pm1)$ by fitting its parameters according to Eq.~(\ref{eq:p_out_stokes}) using the known input states and measured output powers.
We note that each diffraction output intensity is defined by a row of the matrix $\mathbf{B}$ that contains four elements, and therefore four or more calibration measurements are needed to determine uniquely all the $\mathbf{B}$ elements. We intentionally use a large size of the calibration set to reduce the effect of random noise in individual measurements through the averaging.



\begin{figure}
    \centering
    \includegraphics[width=\textwidth]{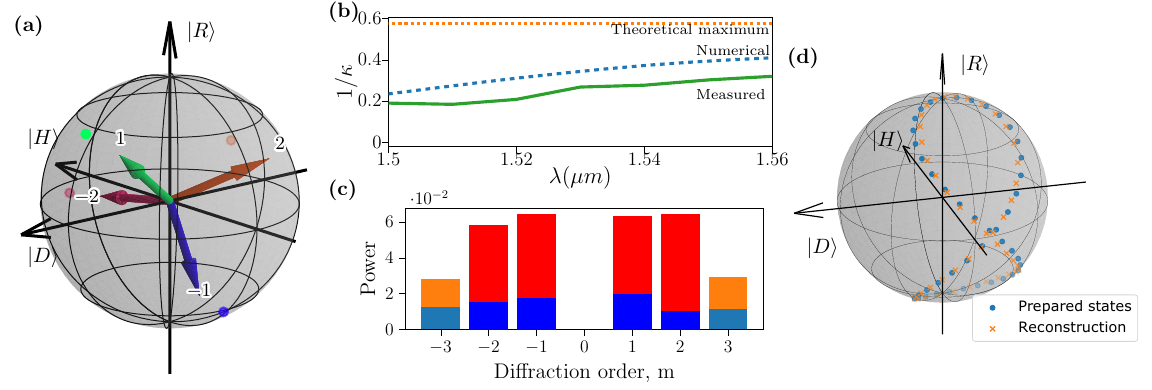}
    \vspace{0.5em}
    \caption{\label{fig:results}
        (a)~Poincar\'e sphere representation of the basis states as calculated from the $(\pm2, \pm1)$ diffraction orders of the fabricated metasurface. These are, as predicted, partially polarized states which deviate from the original numerical design, shown as correspondingly coloured dots on the Poincar\'e sphere.
        (b)~The experimentally characterized inverse condition numbers of the metasurface across a wavelength range are plotted using the solid green line. The blue, dashed line represents the numerical result through CST Studio simulations of the optimal metasurface design, and the orange, dotted line represents the theoretical maximum.
        (c)~The minimum (blue) and maximum (red) power directed to each diffraction order, as determined by experiment. Diffraction orders $\pm3$ were measured but not utilized in the calculations, and are shown for comparison.
        (d)~Poincar\'e sphere showing a comparison of input states versus the reconstructed states using the experimentally characterized metasurface.
    }
\end{figure}

We compute the polarization bases of the measured instrument matrix
and plot a representative example in Fig.~\ref{fig:results}(a), corresponding to the wavelength of $1560$~nm. Here, we see that the metasurface operation has deviated from the designed values as shown in Fig.~\ref{fig:parameters_and_bases}(a), which may be ascribed to 
fabrication errors. Nevertheless, these do not prevent the polarization characterization as we discuss below.

We use the measured instrument matrix to compute the inverse condition number dependence on the wavelengths, see Fig.~\ref{fig:results}(b). Importantly, it is only slightly lower (by up to 15\%) than the numerical design values indicated with the dashed line, across the whole target operating bandwidth. For comparison, the dotted line marks the theoretical maximum of $1/\sqrt{3} \simeq 0.58$~\cite{Foreman:2015-263901:PRL}. Such experimentally achieved performance is close to the value of $1/2.08 \simeq 0.48$ for an interleaved metasurface~\cite{Wang:2018-1104:SCI}, yet providing a fundamentally better beam quality without distortions in the vertical direction.

We use the experimental data to visualize the minimum and maximum powers, across all possible input polarization states and normalized to the input power, at the selected output ports in Fig.~\ref{fig:results}(c). We note that the ratios of these minimum to maximum powers define the polarization extinction, which appears to be of the order of $10^{-1}$. 
Despite such a low extinction, an accurate state reconstruction is still possible using the developed approach based on the framework of generalized quantum measurements.

We demonstrate the reconstruction of the input states from the measured powers at the selected diffraction orders $(\pm2, \pm1)$ after the metasurface. 
From experimental data, a random set of angles for the half- and quarter-waveplates were chosen. From these angles, and taking into account that the first polarizer was positioned vertically, we thus define the input polarization states $\rho_{in}^{(1)}$ before incidence on the metasurface.
We also independently reconstruct the input states from the measured intensities $\Vec{P}_{out}$ and the experimentally characterized instrument matrix $\mathbf{B}$, as:
\begin{equation}
     \Vec{S} = \mathbf{B}^{-1} \cdot \Vec{P}_{out} ,
\end{equation}
and then determine the input density matrix using Eq.~(\ref{eq:rhoStokes}).
We plot the prepared and reconstructed states on the same Poincar\'e sphere for comparison, as shown in Fig.~\ref{fig:results}(d).

In this plot, we show 44 data points, of which only 4 overlap with the calibration dataset, thereby providing a valid test of the reconstruction accuracy for unknown input states.
We quantify the difference between the normalized input ($\Vec{S}^{(in)}$) and reconstructed ($\Vec{S}^{(rec)}$) Stokes vectors as $\delta = 1 - |\langle \Vec{S}^{(in)} \rvert \Vec{S}^{(rec)} \rangle|^2 / ( \langle\Vec{S}^{(in)} \rvert \Vec{S}^{(in)} \rangle \langle\Vec{S}^{(rec)} \rvert \Vec{S}^{(rec)} \rangle)$, such that $\delta = 0$ only if the vectors are the same, up to a total intensity.
Then, we find that for the states in Fig.~\ref{fig:results}(d), the error in reconstruction is $\delta \le 2.9\%$.


Whereas multi-photon quantum experiments are beyond the scope of the current work, we use the experimentally characterized instrument matrix of the metasurface to calculate an inverse condition number for two-photon reconstruction and find that $\kappa^{-1} \simeq 0.15$ can be achieved at a wavelength of $1520nm$ (see Supplementary Information Section~S2 for a wavelength dependence), close to the theoretically optimized value of $0.21$.


\section*{Conclusions}

We have presented a general approach for the complete measurement of polarization in both quantum multi-photon states and classical light using nanostructured metasurfaces. Our method nontrivially combines the principles of single-shot polarimetry and generalized quantum measurements with POVM formalism, 
enabling accurate polarization reconstruction even in the presence of significant fabrication errors or environmental changes, by performing a simple device calibration.
Our experimental measurements demonstrate a complete reconstruction of classical polarization states with a maximum error of $2.9\%$, while maintaining a high optical beam quality after the metasurface for efficient detection. 
%
We anticipate that our new concept will facilitate diverse 
%
quantum and classical applications, from laboratories to satellite imaging systems, benefiting from extremely compact meta-devices providing real-time polarimetric measurements. 

\section*{Acknowledgments}
We thank Jihua Zhang and Dragomir Neshev for the helpful discussions.

\section*{Funding sources}
This work was supported by the Australian Research Council (NI210100072, CE200100010), US AOARD (19IOA053), the Thuringian Ministry for Economy, Science, and Digital Society (2021 FGI 0043), the German Federal Ministry of Education and Research (FKZ 13N14877), the German Research Foundation (IRTG 2675), the European Union through the ERASMUS+ program, the German Academic Exchange Service (grant 57388353), and UA-DAAD exchange scheme.
This work used the ACT node of the NCRIS-enabled Australian National Fabrication Facility (ANFF-ACT).

\section*{Supporting information}
Table~S1; numerically optimized values of phase shifts corresponding to the metasurface renders shown in Figs.~2(a,c), Fig.~S1; figure plotting experimentally calculated two-photon inverse condition number.




\appendix


\bibliography{art_metagrating} 

\newpage

\begin{center}
\textbf{\Large Supplementary Materials}
\end{center}
\setcounter{equation}{0}
\setcounter{figure}{0}
\setcounter{table}{0}
\setcounter{page}{1}
\makeatletter
\renewcommand{\theequation}{S\arabic{equation}}
\renewcommand{\thefigure}{S\arabic{figure}}
\renewcommand{\thetable}{S\arabic{table}}
\renewcommand{\bibnumfmt}[1]{[S#1]}
\renewcommand{\citenumfont}[1]{S#1}

\section{Optimized metasurface parameters} \label{sec:MSparameters}

Table~\ref{table:parameters} contains the numerically optimized values of phase shifts along the $x$ and $y$ directions as well as the angle of rotations of each nanoresonator, corresponding to the metasurface renders shown in Figs.~2(a,c).

\begin{table}
    \centering
    \begin{tabular}{|c||ccc||ccc|}
        \hline
        \multicolumn{1}{|c||}{Resonator no.} & \multicolumn{3}{c||}{One-photon polarimetry}  & \multicolumn{3}{c|}{Two-photon polarimetry}   \\
         & $\phi_x$ & $\phi_y$ & $\theta$       & $\phi_x$ & $\phi_y$ & $\theta$       \\
        \hline
         1 & -0.29 &  -0.15 &    0.43 &  -1.19  & -1.5   &   0.72 \\
         2 & -0.45 &   0.11 &   -1.34 &  -1.08  &  0.85  &   0.94 \\
         3 & -0.73 &   0.33 &   -1.48 &  -1.09  &  0.078 &  -0.092 \\
         4 & -1.48 &  -0.30 &    1.1  &  -1.31  & -1.12  &  -0.28 \\
         5 & -1.06 &  -0.22 &    0.86 &  -1.39  &  1.44  &   0.76 \\
         6 &  1.08 &  -0.38 &    0.75 &   1.11  &  1.38  &   0.53 \\
         7 & -1.51 &  -0.42 &    0.83 &   1.47  & -1.42  &  -1.29 \\
         8 &  1.52 &  -0.71 &    0.86 &  -1.4   &  1.39  &   0.128 \\
         9 &  1.36 &  -0.72 &    1.01 &  -0.40  &  1.19  &  -0.0016 \\
        10 &  1.54 &  -0.26 &    1.0  &   0.7   &  1.22  &  -0.362 \\
        11 &  1.18 &  -1.13 &   -1.25 &   1.06  &  0.072 &   0.26 \\
        12 &  1.05 &  -0.48 &    1.52 &  -0.77  &  1.11  &   1.42 \\
        13 &  0.29 &  -0.41 &    1.56 &   1.2   &  1.52  &  -0.29 \\
        14 & -0.76 &  -0.34 &    1.37 &   1.46  &  1.34  &   1.17 \\
        15 & -1.22 &  -0.41 &    1.02 &  -1.55  & -1.07  &  -0.72 \\
        16 & -1.29 &  -0.33 &    0.77 &  0.288  & -1.25  &   0.66 \\
        \hline
   \end{tabular}
    \caption{Phases and rotations in radians of nanoresonators forming metasurfaces optimized for one- and two-photon polarimetry.}
    \label{table:parameters}
\end{table}

\pagebreak

\section{2-photon condition number}
\begin{figure}
    \includegraphics[width=0.9\linewidth]{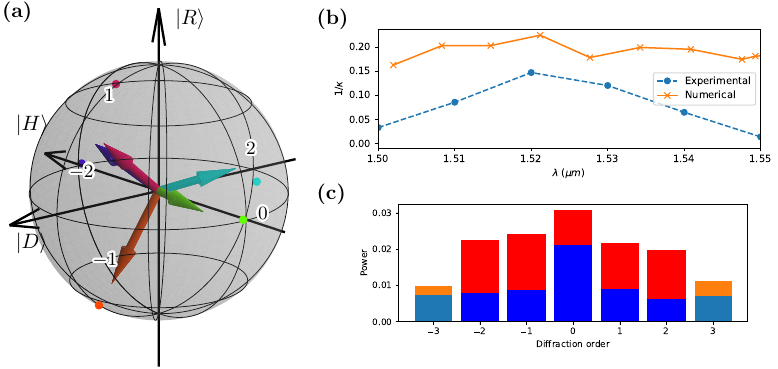}
    \caption{
    (a)~Poincar\'e sphere representation of the basis states as calculated from the $(\pm2, \pm1, 0)$ diffraction orders of the fabricated metasurface.
    (b)~The two-photon state reconstruction condition numbers vs. the wavelength determined from the experimentally characterized metasurface instrument matrix (green line) and from the finite-difference numerical simulations with CST studio (blue line).
    (c)~The minimum (blue) and maximum (red) power directed to each diffraction order, as determined by experiment.
    }
\end{figure}

\end{document}